\documentclass[3p,12pt]{elsarticle}

\usepackage{amssymb,latexsym,amsmath,multirow,url}

\journal{}

\newcommand{\set}[1]{\left\{#1\right\}}

\newcommand{\p}{\partial}

\newtheorem{thm}{Theorem}[section]

\newtheorem{lem}[thm]{Lemma}

\newtheorem{Rem}[thm]{Remark}
\newproof{pf}{Proof}

\begin{document}

\begin{frontmatter}



\title{Fast shape reconstruction of perfectly conducting cracks by using a multi-frequency topological derivative strategy}


\author{Won-Kwang Park}
\ead{parkwk@kookmin.ac.kr}
\address{Department of Mathematics, Kookmin University, Seoul, 136-702, Korea.}

\begin{abstract}
This paper concerns a fast, one-step iterative technique of imaging extended perfectly conducting cracks with Dirichlet boundary condition. In order to reconstruct the shape of cracks from scattered field data measured at the boundary, we introduce a topological derivative-based electromagnetic imaging function operated at several nonzero frequencies. The properties of the imaging function are carefully analyzed for the configurations of both symmetric and non-symmetric incident field directions. This analysis explains why the application of incident fields with symmetric direction operated at multiple frequencies guarantees a successful reconstruction. Various numerical simulations with noise-corrupted data are conducted to assess the performance, effectiveness, robustness, and limitations of the proposed technique.
\end{abstract}


\end{frontmatter}





\section{Introduction}
One goal of the inverse scattering problem is to find the shape of bulk or flaw in a medium using the scattered field data measured at the boundary. This is an old but it has several of applications such as detection of cracks in material engineering, ultrasound imaging in medical sciences, and scanning anti-personnel mines hidden in the ground in military services. Related researches may be found in \cite{A1,AIL1,D,DLL,DL,S} and the references therein. In many studies, various shape reconstruction algorithms have been developed, most of which are focused on the minimization of the least-square functional by using a Newton-type iteration strategy, e.g., level-set method \cite{ADIM,DL,PL4}. The merits of the iteration strategy is that it does not require a very large number of boundary measurements for complete shape reconstruction; instead, in order to guarantee successful reconstruction, it demands very high computational expenditures, optimal regularization terms related to the problems at hand, calculation of a complex Fr{\'e}chet derivative at each iteration, and \textit{a priori} information of the unknown target to be reconstructed. Nevertheless, even if these conditions are satisfied, the iteration strategy must begin with an initial guess that is close to the true one in order to avoid undesirable situations such as non-convergence or falling into the local minima. Therefore, development of a mathematical theory and an algorithm for generating a good initial guess is an important research topic.

Conversely, certain non-iterative shape reconstruction algorithms have been proposed; for example, MUltiple SIgnal Classification (MUSIC) algorithm \cite{AGKPS,AKLP,PL1,PL3}, linear sampling method \cite{C,KR}, Kirchhoff migration \cite{AGKPS,P3,P4,PL2}, and Fourier inversion based one \cite{AIL2,AK,AMV}. In contrast with the iterative strategy, these algorithms require a large amount of incident field data and a significant number of boundary measurements. However, if these conditions are satisfied, non-iterative shape reconstruction algorithms can prove to be effective, fast, and easy to extend to multiple targets.

Topological derivative concept is a non-iterative strategy. This concept was originally developed for the shape optimization problem, but its application to rapid shape reconstruction has only recently been proven. Related works can be found in \cite{AGJK,AKLP,AM,B,CR,MKP,NB,P2,P5,P6,SZ} and references therein. One of the advantages of topological derivative concept is that it does not require a large amount of many incident field data; however, a reduction in the amount of this data has been reported to result in poor resolution of the reconstructed shape.

The purpose of this literature is to establish an effective and fast reconstruction algorithm for detecting both the location and shape of perfectly conducting crack(s) from boundary-measured scattered field data by using only a small amount of incident field data. Based on the structure of topological derivative imaging functional operated at a fixed-frequency \cite{MKP} and multi-frequency imaging schemes \cite{AGKPS,P1}, we introduce a multi-frequency topological derivative concept. Note that recent work \cite[Formula (1.33)]{A2} has briefly introduced a multi-frequency topological derivative but through a more detailed analysis, we investigate the structure of the proposed multi-frequency topological derivative and its benefits over the traditional topological derivative.

The remainder of this paper is organized as follows. Section \ref{Sec2} briefly introduces the traditional topological derivative by creating a linear, narrow crack. Section \ref{Sec3} describes the design of a multi-frequency topological derivative imaging function and analyzes it to investigate its properties. Section \ref{Sec4} presents various numerical experiments to assess the performance of the proposed imaging function under various situations. Section \ref{Sec5} contains a short conclusion and some remarks on future work.

\section{Topological derivative: an inspection}\label{Sec2}
Assume that a perfectly conducting crack $\mathcal{C}$ is completely hidden in a homogeneous domain $\mathcal{D}\subset\mathbb{R}^2$ with a smooth boundary $\mathcal{B}$. We denote $\vec{x}_c$ as a two-dimensional vector that lies on crack $\mathcal{C}$.

Throughout this paper, we consider the so-called Transverse Magnetic (TM) polarization case, letting $u_m^{\mathrm{total}}(\vec{x};k)$ be the (single-component) electric field that satisfies the following boundary value problem:
\begin{equation}\label{Helmholtz}
\left\{\begin{array}{rcl}
\Delta u_m^{\mathrm{total}}(\vec{x};k)+k^2 u_m^{\mathrm{total}}(\vec{x};k)=0&\mbox{in}&\mathcal{D}\backslash\overline{\mathcal{C}}\\
\noalign{\medskip}u_m^{\mathrm{total}}(\vec{x};k)=0&\mbox{on}&\mathcal{C}\\
\noalign{\medskip}\displaystyle\frac{\p u_m^{\mathrm{total}}(\vec{x};k)}{\p\vec{\nu}(\vec{x})}=\frac{\p\exp(ik\vec{\theta}_m\cdot\vec{x})}{\p\vec{\nu}(\vec{x})}&\mbox{on}&\mathcal{B},
\end{array}\right.
\end{equation}
where $\vec{\nu}(\vec{x})$ represents the unit outward normal to $\vec{x}\in\mathcal{B}$ and $\vec{\theta}_m$, $m=1,2,\cdots,M$, denotes the two-dimensional vector on the unit circle $\mathcal{S}^1$. Throughout this paper, we let $k$ denote a strictly positive wavenumber and presume $k^2$ to not be an eigenvalue of (\ref{Helmholtz}). Similarly, let $u_m^{\mathrm{back}}(\vec{x};k)=\exp(ik\vec{\theta}_m\cdot\vec{x})$ be the solution of equation (\ref{Helmholtz}) without $\mathcal{C}$. Then, the problem we consider here is the computation of the topological derivative of the residual function depending on the solution $u_m^{\mathrm{total}}(\vec{x};k)$:
\begin{equation}\label{Energy}
  \mathbb{R}(\mathcal{D};k):=\frac12\sum_{m=1}^{M}\left|\left|u_m^{\mathrm{total}}(\vec{x};k)-u_m^{\mathrm{ back}}(\vec{x};k)\right|\right|_{L^2(\mathcal{B})}^2=\frac12\sum_{m=1}^{M}\int_{\mathcal{B}}\left|u_m^{\mathrm{total}}(\vec{x};k)-u_m^{\mathrm{ back}}(\vec{x};k)\right|^2d\mathcal{B}(\vec{x}),
\end{equation}
where $|u_m^{\mathrm{total}}(\vec{x};k)|^2=u_m^{\mathrm{total}}(\vec{x};k)\overline{u_m^{\mathrm{total}}(\vec{x};k)}$.

In order to compute the topological derivative, let us create a small linear crack $\mathcal{L}$ of length $2\ell(\ll k)$ centered at $\vec{x}_s=(x_s^1,x_s^2)\in\mathcal{D}\backslash\mathcal{B}$ such that
\[\mathcal{L}=\left\{(\xi,x_s^2):x_s^1-\ell\leq\xi\leq x_s^1+\ell\right\},\]
and denote $\mathcal{D}\vee\mathcal{L}$ as that domain. Then, because of the change in the topology of $\mathcal{D}$, we can consider the topological derivative $\mathbb{R}_{\mathrm{TD}}(\vec{x}_s;k)$ based on the residual function $\mathbb{R}(\mathcal{D};k)$ with respect to point $\vec{x}_s$ as follows:
\begin{equation}\label{TopDerivative}
  \mathbb{R}_{\mathrm{TD}}(\vec{x}_s;k)=\lim_{\ell\to0+}\frac{\mathbb{R}(\mathcal{D}\vee\mathcal{L};k)-\mathbb{R}(\mathcal{D};k)}{\phi(\ell;k)},
\end{equation}
where $\phi(\ell;k)\longrightarrow0$ as $\ell\longrightarrow0+$. From relationship (\ref{TopDerivative}), we have the following asymptotic expansion:
\[\mathbb{R}(\mathcal{D}\vee\mathcal{L};k)=\mathbb{R}(\mathcal{D};k)+\phi(\ell;k)
\mathbb{R}_{\mathrm{TD}}(\vec{x}_s;k)+o(\phi(\ell;k)).\]

Then, the topological derivative $\mathbb{R}_{\mathrm{TD}}(\vec{x}_s;k)$ with $M$ different incident waves at a given wavenumber $k$ is as follows (see \cite{MKP} for its derivation).

\begin{lem}[Topological derivative]\label{ThTop}
  Let $\Re(f)$ denote the real part of $f$. Then, the topological derivative corresponding to (\ref{Energy}) is given by
  \begin{equation}\label{Topderivative2}
    \mathbb{R}_{\mathrm{TD}}(\vec{x}_s;k)=\Re\left[\sum_{m=1}^{M}v_m^{\mathrm{ adjnt}}(\vec{x}_s;k)\overline{u_m^{\mathrm{back}}(\vec{x}_s;k)}\right],
  \end{equation}
  where $v_m^{\mathrm{adjnt}}(\cdot;k)$ satisfies the following adjoint problem:
  \begin{equation}\label{Adjoint}
    \left\{\begin{array}{rcl}
      \displaystyle\Delta v_m^{\mathrm{adjnt}}(\vec{x};k)+k^2v_m^{\mathrm{ adjnt}}(\vec{x};k)=0&\mbox{in}&\mathcal{D}\\
      \noalign{\medskip}\displaystyle\frac{\partial v_m^{\mathrm{ adjnt}}(\vec{x};k)}{\partial\vec{\nu}(\vec{x})}=u_m^{\mathrm{total}}(\vec{x};k)-u_m^{\mathrm{ back}}(\vec{x};k)&\mbox{on}&\mathcal{B}
    \end{array}\right.
  \end{equation}
\end{lem}

\section{Multi-frequency topological derivative: introduction and analysis}\label{Sec3}
In this section, we introduce a multi-frequency topological derivative imaging function and explore its properties. Based on recent research \cite{AGJK,AKLP,P5,P6}, a sufficiently large number of incident directions are required to obtain a good result. This means that if $M$ is small, an image with poor resolution will be reconstructed. Moreover, when boundary measurement data are affected by a large amount of random noise, (\ref{Topderivative2}) will fail to reconstruct a good image. Inspired by multi-frequency imaging algorithms \cite{AGKPS,P3,P4,PL2}, we consider the following multi-frequency topological derivative for several wavenumbers $\{k_f:f=1,2,\cdots,F\}$:
\begin{equation}\label{MultiFrequencyTopologicalDerivative}
  \mathbb{R}_{\mathrm{MTD}}(\vec{x}_s;F)=\frac{1}{F}\sum_{f=1}^{F} \frac{\mathbb{R}_{\mathrm{TD}}(\vec{x}_s;k_f)}{\displaystyle\max_{\vec{x}_s\in\mathcal{D}}|\mathbb{R}_{\mathrm{TD}}(\vec{x}_s;k_f)|},
\end{equation}
where $\mathbb{R}_{\mathrm{TD}}(\vec{x}_s;k_f)$ is given by (\ref{Topderivative2}) for $k=k_f$, $f=1,2,\cdots,F$.

In this section, we investigate the structure of (\ref{MultiFrequencyTopologicalDerivative}). For this purpose, we recall a useful result (see \cite{MKP}) as follows.
\begin{lem}\label{lemMKP}
  Let $A\sim B$ imply that there exists a constant $C$ such that $A=BC$. Then, for $\vec{x}_s\in\mathcal{D}$ and $\vec{x}_c\in\mathcal{C}$, (\ref{Topderivative2}) satisfies
  \begin{equation}\label{Topderivative3}
    \mathbb{R}_{\mathrm{TD}}(\vec{x}_s;k)\sim\Re\left[\frac{i}{k}\sum_{m=1}^{M}\exp(ik\vec{\theta}_m\cdot(\vec{x}_c-\vec{x}_s))\Phi(\vec{x}_c,\vec{x}_s;k)\right].
  \end{equation}
Here $\Phi(x,y;k)$ denotes the two-dimensional time-harmonic fundamental solution (or Green's function) of Helmholtz equation
\[\Phi(x,y;k)=-\frac{i}{4}H_0^1(k|x-y|)=-\frac{i}{4}\bigg(J_0(k|x-y|)+iY_0(k|x-y|)\bigg),\]
where $H_0^1$, $J_0$, and $Y_0$ denote the Hankel function of the first kind, Bessel function, and Neumann function of order zero, respectively.
\end{lem}

Consequently, (\ref{MultiFrequencyTopologicalDerivative}) is analyzed for the following three cases of interest.

\begin{thm}[Case 1: Symmetric incident directions]\label{TheoremSymmetric}
  Under the configuration of a small number of symmetric incident directions, the structure of (\ref{MultiFrequencyTopologicalDerivative}) is
  \[\mathbb{R}_{\mathrm{MTD}}(\vec{x}_s;F)\sim\sum_{m=1}^{M}\frac{1}{\displaystyle |\vec{x}_c-\vec{x}_s|\sqrt{1-\bigg(\vec{\theta}_m\cdot\frac{\vec{x}_c-\vec{x}_s}{|\vec{x}_c-\vec{x}_s|}\bigg)^2}},\]
  for a sufficiently large $F$ and $k_F\longrightarrow\infty$.
\end{thm}
\begin{pf} Assume that the symmetric incident directions are of the form:
\[\vec{\theta}_m:=\bigg(\cos\frac{2(m-1)\pi}{M},\sin\frac{2(m-1)\pi}{M}\bigg),\quad m=1,2,\cdots,M,\]
where $M$ is an even number, say, $M=2N$. Based on this form, $\vec{\theta}_l$ satisfies $\vec{\theta}_l=-\vec{\theta}_{l+N}$ for $l=1,2,\cdots,N$.

In contrast with Lemma \ref{lemMKP}, we assume that $M$ is not sufficiently large. Let $\Im(f)$ denote the imaginary part of $f$ then since
\begin{align*}
  \Re\bigg[\exp(ik\vec{\theta}_l\cdot(\vec{x}_c-\vec{x}_s))\bigg]&=\Re\bigg[\exp(ik\vec{\theta}_{l+N}\cdot(\vec{x}_c-\vec{x}_s))\bigg]\\
  \Im\bigg[\exp(ik\vec{\theta}_l\cdot(\vec{x}_c-\vec{x}_s))\bigg]&=-\Im\bigg[\exp(ik\vec{\theta}_{l+N}\cdot(\vec{x}_c-\vec{x}_s))\bigg],
\end{align*}
(\ref{Topderivative3}) becomes
\begin{align*}
  \mathbb{R}_{\mathrm{TD}}(\vec{x}_s;k_f)&\sim\Re\left[\frac{i}{k_f}\sum_{m=1}^{M}\exp(ik_f\vec{\theta}_m\cdot(\vec{x}_c-\vec{x}_s))\Phi(\vec{x}_c,\vec{x}_s;k_f)\right]\\
  &=\Re\left[\frac{1}{2k_f}\sum_{m=1}^{N}\cos(k_f\vec{\theta}_m\cdot(\vec{x}_c-\vec{x}_s))\{J_0(k_f|\vec{x}_c-\vec{x}_s|)+iY_0(k_f|\vec{x}_c-\vec{x}_s|)\}\right]\\
  &=\frac{1}{2k_f}\sum_{m=1}^{N}\cos(k_f\vec{\theta}_m\cdot(\vec{x}_c-\vec{x}_s))J_0(k_f|\vec{x}_c-\vec{x}_s|).
\end{align*}
Therefore,
\[\frac{\mathbb{R}_{\mathrm{TD}}(\vec{x}_s;k_f)}{\displaystyle\max_{\vec{x}_s\in\mathcal{D}}|\mathbb{R}_{\mathrm{TD}}(\vec{x}_s;k_f)|}\approx\frac{1}{N}\sum_{m=1}^{N}\cos(k_f\vec{\theta}_m\cdot(\vec{x}_c-\vec{x}_s))J_0(k_f|\vec{x}_c-\vec{x}_s|)\]

Let us recall that for $\Re(\mu)>-1$ (see \cite[formula 11.4.37 (p. 487)]{AS})
\[\int_0^{\infty}\cos at J_\mu(bt)dt=\left\{\begin{array}{ccc}
  \medskip\displaystyle \frac{1}{\sqrt{b^2-a^2}}\cos\bigg(\mu\sin^{-1}\frac{a}{b}\bigg)&\mbox{if}&0\leq b<a\\
  \displaystyle -\frac{b^\mu}{\sqrt{a^2-b^2}(a+\sqrt{a^2-b^2})^\mu}\cos\bigg(\frac{\mu\pi}{2}\bigg)&\mbox{if}&0<a<b.
\end{array}\right.\]
Let $a=\vec{\theta}_m\cdot(\vec{x}_c-\vec{x}_s)$, $b=|\vec{x}_c-\vec{x}_s|$, and $\mu=0$. Then since
\begin{equation}\label{Inequalityab}
  b^2-a^2=|\vec{x}_c-\vec{x}_s|^2\bigg[1-\bigg(\vec{\theta}_m\cdot\frac{\vec{x}_c-\vec{x}_s}{|\vec{x}_c-\vec{x}_s|}\bigg)^2\bigg]\geq0,
\end{equation}
by letting $k_F\longrightarrow\infty$, (\ref{MultiFrequencyTopologicalDerivative}) becomes
\begin{align}
\begin{aligned}\label{StructureMultiTopDerivative1}
  \mathbb{R}_{\mathrm{MTD}}(\vec{x}_s;F)&=\frac{1}{F}\sum_{f=1}^{F}\frac{\mathbb{R}_{\mathrm{TD}}(\vec{x}_s;k_f)}{\displaystyle\max_{\vec{x}_s\in\mathcal{D}}|\mathbb{R}_{\mathrm{TD}}(\vec{x}_s;k_f)|} \approx\frac{1}{N}\sum_{m=1}^{N}\int_0^{\infty}\cos(k\vec{\theta}_m\cdot(\vec{x}_c-\vec{x}_s))J_0(k|\vec{x}_c-\vec{x}_s|)dk\\
  &=\frac{1}{N}\sum_{m=1}^{N}\frac{1}{\displaystyle |\vec{x}_c-\vec{x}_s|\sqrt{1-\bigg(\vec{\theta}_m\cdot\frac{\vec{x}_c-\vec{x}_s}{|\vec{x}_c-\vec{x}_s|}\bigg)^2}}.
\end{aligned}
\end{align}
\end{pf}

\begin{thm}[Case 2: Non-symmetric incident directions]\label{TheoremNonSymmetric}
  Under the configuration of a small number of non-symmetric incident directions, the structure of (\ref{MultiFrequencyTopologicalDerivative}) becomes
  \[\mathbb{R}_{\mathrm{MTD}}(\vec{x}_s;F)\sim\sum_{m=1}^{M}\frac{\displaystyle 1-\frac{2}{\pi}\sin^{-1}\bigg(\vec{\theta}_m\cdot\frac{\vec{x}_c-\vec{x}_s}{|\vec{x}_c-\vec{x}_s|}\bigg)}{\displaystyle |\vec{x}_c-\vec{x}_s|\sqrt{1-\bigg(\vec{\theta}_m\cdot\frac{\vec{x}_c-\vec{x}_s}{|\vec{x}_c-\vec{x}_s|}\bigg)^2}},\]
  for a sufficiently large $F$ and $k_F\longrightarrow\infty$.
\end{thm}
\begin{pf} In constrast with Theorem \ref{TheoremSymmetric}, we assume that the incident directions are non-symmetric. Then, (\ref{Topderivative3}) can be written as
\begin{align*}
  \mathbb{R}_{\mathrm{TD}}(\vec{x}_s;k_f)\sim&\Re\left[\frac{i}{k_f}\sum_{m=1}^{M}\exp(ik_f\vec{\theta}_m\cdot(\vec{x}_c-\vec{x}_s)) \Phi(\vec{x}_c,\vec{x}_s;k_f)\right]\\
  =&\Re\left[\frac{1}{4k_f}\sum_{m=1}^{M}\exp(ik_f\vec{\theta}_m\cdot(\vec{x}_c-\vec{x}_s)) \{J_0(k_f|\vec{x}_c-\vec{x}_s|)+iY_0(k_f|\vec{x}_c-\vec{x}_s|)\}\right]\\
  =&\frac{1}{4k_f}\sum_{m=1}^{M}\bigg[\cos(k_f\vec{\theta}_m\cdot(\vec{x}_c-\vec{x}_s))J_0(k_f|\vec{x}_c-\vec{x}_s|)\\ &-\sin(k_f\vec{\theta}_m\cdot(\vec{x}_c-\vec{x}_s))Y_0(k_f|\vec{x}_c-\vec{x}_s|)\bigg],
\end{align*}
and therefore
\begin{align*}
  \frac{\mathbb{R}_{\mathrm{TD}}(\vec{x}_s;k_f)}{\displaystyle\max_{\vec{x}_s\in\mathcal{D}}|\mathbb{R}_{\mathrm{TD}}(\vec{x}_s;k_f)|}\approx\frac{1}{2M}\sum_{m=1}^{M} \bigg[&\cos(k_f\vec{\theta}_m\cdot(\vec{x}_c-\vec{x}_s))J_0(k_f|\vec{x}_c-\vec{x}_s|)\\ &-\sin(k_f\vec{\theta}_m\cdot(\vec{x}_c-\vec{x}_s))Y_0(k_f|\vec{x}_c-\vec{x}_s|)\bigg].
\end{align*}
Let us recall an integral formula in \cite[formula 11.4.40 (p. 487)]{AS},
\[\int_0^{\infty}\exp(iat)Y_0(bt)dt=\left\{
\begin{array}{ccc}
  \medskip\displaystyle\frac{2i}{\pi\sqrt{b^2-a^2}}\sin^{-1}\left(\frac{a}{b}\right)&\mbox{if}&0\leq a<b\\
  \displaystyle\frac{2i}{\pi\sqrt{a^2-b^2}}\ln\bigg(\frac{a-\sqrt{a^2-b^2}}{b}\bigg)-\frac{1}{\sqrt{a^2-b^2}}&\mbox{if}&0<b<a.
\end{array}\right.\]

Then, if $k_F\longrightarrow\infty$, we can evaluate
\begin{multline}\label{ApproximationSinExp}
  \sum_{f=1}^{F}\sin(k_f\vec{\theta}_m\cdot(\vec{x}_c-\vec{x}_s))Y_0(k_f|\vec{x}_c-\vec{x}_s|)\approx
  \int_0^\infty\sin(k\vec{\theta}_m\cdot(\vec{x}_c-\vec{x}_s))Y_0(k|\vec{x}_c-\vec{x}_s|)dk\\ =\Im\int_0^\infty\exp(k\vec{\theta}_m\cdot(\vec{x}_c-\vec{x}_s))Y_0(k|\vec{x}_c-\vec{x}_s|)dk
  =\frac{\displaystyle\frac{2}{\pi}\sin^{-1}\bigg(\vec{\theta}_m\cdot\frac{\vec{x}_c-\vec{x}_s}{|\vec{x}_c-\vec{x}_s|}\bigg)}{\displaystyle |\vec{x}_c-\vec{x}_s|\sqrt{1-\bigg(\vec{\theta}_m\cdot\frac{\vec{x}_c-\vec{x}_s}{|\vec{x}_c-\vec{x}_s|}\bigg)^2}}.
\end{multline}

Hence, by combining (\ref{Inequalityab}), (\ref{StructureMultiTopDerivative1}), and (\ref{ApproximationSinExp}), we can obtain
\begin{equation}\label{StructureMultiTopDerivative2}
  \mathbb{R}_{\mathrm{MTD}}(\vec{x}_s;F)=\frac{1}{F}\sum_{f=1}^{F}\frac{\mathbb{R}_{\mathrm{TD}}(\vec{x}_s;k_f)}{\displaystyle\max_{\vec{x}_s\in\mathcal{D}}|\mathbb{R}_{\mathrm{TD}}(\vec{x}_s;k_f)|} \approx\frac{1}{2M}\sum_{m=1}^{M}\frac{\displaystyle 1-\frac{2}{\pi}\sin^{-1}\bigg(\vec{\theta}_m\cdot\frac{\vec{x}_c-\vec{x}_s}{|\vec{x}_c-\vec{x}_s|}\bigg)}{\displaystyle |\vec{x}_c-\vec{x}_s|\sqrt{1-\bigg(\vec{\theta}_m\cdot\frac{\vec{x}_c-\vec{x}_s}{|\vec{x}_c-\vec{x}_s|}\bigg)^2}}.
\end{equation}
\end{pf}

\begin{thm}[Case 3: Sufficiently large number of incident directions]\label{TheoremManyDirection}
  Under the configuration of a sufficiently large number of incident directions, the structure of (\ref{MultiFrequencyTopologicalDerivative}) becomes
  \[\mathbb{R}_{\mathrm{MTD}}(\vec{x}_s;F)\approx\left\{
  \begin{array}{ccc}
    \Lambda(|\vec{x}_c-\vec{x}_s|;k_F)-\Lambda(|\vec{x}_c-\vec{x}_s|;k_1)
     & \mbox{if} & k_F<+\infty,\\
    \noalign{\medskip}\delta(\vec{x}_c,\vec{x}_s) & \mbox{if} & k_F\longrightarrow\infty,
  \end{array}
  \right.\]
  for an adequately large $F$. Here, $\Lambda(x)$ is defined as
  \begin{equation}\label{FunctionLambda}
    \Lambda(x;k):=k\bigg(J_0(kx)^2+J_1(kx)^2\bigg),
  \end{equation}
  and $\delta(\vec{x}_c,\vec{x}_s)$ denotes the Dirac delta function.
\end{thm}
\begin{pf}Suppose that $M$ is large enough, i.e., $M\longrightarrow\infty$, then by \cite[Lemma 4.1]{G}, (\ref{Topderivative3}) can be approximated as
  \begin{align*}
    \mathbb{R}_{\mathrm{TD}}(\vec{x}_s;k)&\sim\Re\left[\frac{i}{k}\sum_{m=1}^{M} \exp(ik\vec{\theta}_m\cdot(\vec{x}_c-\vec{x}_s))\Phi(\vec{x}_c,\vec{x}_s;k)\right]\\
    &\approx\Re\left[\frac{1}{4k}\bigg(J_0(k|\vec{x}_c-\vec{x}_s|)+iY_0(k|\vec{x}_c-\vec{x}_s|)\bigg) \int_{\mathcal{S}^1}\exp(ik\vec{\theta}\cdot(\vec{x}_c-\vec{x}_s))d\vec{\theta}\right]\\
    &=\frac{\pi}{2k}J_0(k|\vec{x}_c-\vec{x}_s|)^2.
  \end{align*}

  Assume that $F$ is large enough and that $k_F<+\infty$. Then applying an integral formula of the Bessel function \cite[page 35]{R}
  \[\int J_0(t)^2dt=t\bigg(J_0(t)^2+J_1(t)^2\bigg)+\int J_1(t)^2dt\]
  and a change of variable $t=k|\vec{x}_c-\vec{x}_s|$ yields
  \begin{align}
  \begin{aligned}\label{StructureMultiTopDerivative3}
    \sum_{f=1}^{F}\mathbb{R}_{\mathrm{TD}}(\vec{x}_s;k_f)&\sim\frac{1}{k_F-K_1}\int_{k_1}^{k_F} J_0(k|\vec{x}_c-\vec{x}_s|)^2dk=\frac{1}{|\vec{x}_c-\vec{x}_s|}\int_{k_1|\vec{x}_c-\vec{x}_s|}^{k_F|\vec{x}_c-\vec{x}_s|}J_0(t)^2dt\\
    =&\Lambda(|\vec{x}_c-\vec{x}_s|;k_F)-\Lambda(|\vec{x}_c-\vec{x}_s|;k_1)+\int_{k_1}^{k_F}J_1(k|\vec{x}_c-\vec{x}_s|)^2dk,
    \end{aligned}
  \end{align}
  where $\Lambda(t;k)$ is given in (\ref{FunctionLambda}). Note that the term
  \[\Theta(|\vec{x}_c-\vec{x}_s|,k):=\int_{k_1}^{k_F}J_1(k|\vec{x}_c-\vec{x}_s|)^2dk\]
  can be negligible since $\Lambda(|\vec{x}_s-\vec{x}_c|,k_F)=O(k_F)$ and $\Theta(|\vec{x}_c-\vec{x}_s|,k)\ll O(k_F)$, refer to \cite{JKHP}.

  Finally, by applying $k_F\longrightarrow\infty$ in (\ref{StructureMultiTopDerivative3}), we can immediately obtain the following result:
  \begin{equation}\label{StructureMultiTopDerivative4}
    \sum_{f=1}^{F}\mathbb{R}_{\mathrm{TD}}(\vec{x}_s;k_f)\sim\int_0^\infty J_0(k|\vec{x}_c-\vec{x}_s|)^2dk=\left\{
    \begin{array}{ccc}
    +\infty & \mbox{if} & \vec{x}_c=\vec{x}_s \\
    0 & \mbox{if} & \vec{x}_c\ne\vec{x}_s
    \end{array}\right.
  \end{equation}
\end{pf}

\begin{figure}[!ht]
\begin{center}
\includegraphics[width=0.55\textwidth,keepaspectratio=true,angle=0]{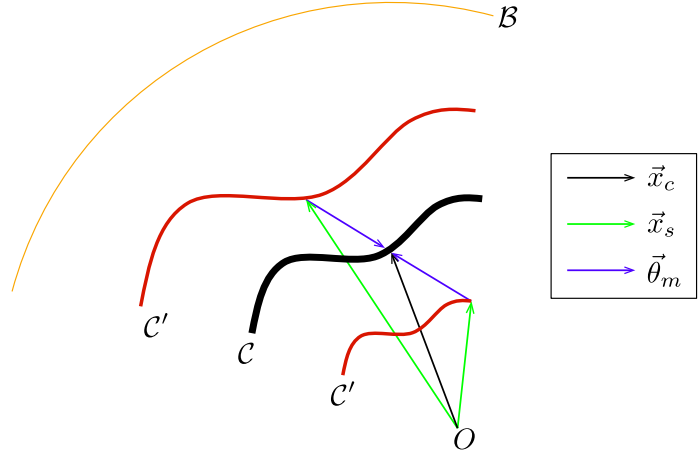}
\caption{\label{RemarkVector}Illustration of Remark \ref{Remark1}. At the location $\vec{x}_s$ such that $\vec{x}_s=\vec{x}_c$ (black colored vector) and $\vec{\theta}_m\cdot\frac{\vec{x}_c-\vec{x}_s}{|\vec{x}_c-\vec{x}_s|}=\pm1$ (green colored vector), true shape of $\mathcal{C}$ and ghost replicas $\mathcal{C}'$ will appear, respectively.}
\end{center}
\end{figure}

\begin{Rem}\label{Remark1}Based on structures (\ref{StructureMultiTopDerivative1}), (\ref{StructureMultiTopDerivative2}), and (\ref{StructureMultiTopDerivative3}), we can summarize some properties of (\ref{MultiFrequencyTopologicalDerivative}) as follows:
\begin{enumerate}\renewcommand{\theenumi}{(P\arabic{enumi})}
  \item\label{P1Remark} Formulas (\ref{StructureMultiTopDerivative1}) and (\ref{StructureMultiTopDerivative2}) reach their maximum value at $\vec{x}_s\in\mathcal{D}$ that satisfies
      \[|\vec{x}_c-\vec{x}_s|=0.\]
      Therefore, at $\vec{x}_s=\vec{x}_c\in\mathcal{C}$, the map of (\ref{MultiFrequencyTopologicalDerivative}) will yield $1$ that the location of $\mathcal{C}$ can be accurately determined. However, at point $\vec{x}_s\in\mathcal{D}$ such that
      \[\vec{\theta}_m\cdot\frac{\vec{x}_c-\vec{x}_s}{|\vec{x}_c-\vec{x}_s|}=\pm1,\]
      the map of (\ref{MultiFrequencyTopologicalDerivative}) will show a large magnitude such that unexpected replicas $\mathcal{C}'$ will appear (see Figure \ref{RemarkVector}).
  \item\label{P2Remark} If $\vec{x}_s\in\mathcal{D}$ satisfies
  \[\frac{|\vec{x}_c-\vec{x}_s|}{\vec{\theta}_m\cdot(\vec{x}_c-\vec{x}_s)}\approx1,\]
  then both the numerator and denominator of (\ref{StructureMultiTopDerivative2}) become $0$. Hence, at this point, (\ref{StructureMultiTopDerivative2}) will yield a large value at not only $\vec{x}_c$ but also other locations in $\mathcal{D}$. Therefore, the configuration of symmetric incident directions will yield a better imaging result (see Figure \ref{GammaNonSymmetric1}).
  \item\label{P3Remark} If the number of incident directions is increased to as many as possible, i.e., $M\longrightarrow\infty$, it is clear that a more accurate location of $\vec{x}_c\in\mathcal{C}$ can be obtained. In this case, an odd number of $M$ will not affect the imaging performance significantly.
  \item\label{P4Remark} Application of multiple frequencies enhances the imaging performance.
\end{enumerate}
\end{Rem}

\section{Numerical experiments}\label{Sec4}
In this section, we exhibit some numerical examples to demonstrate the effectiveness of the proposed algorithm. Throughout this section, the homogeneous domain $\mathcal{D}$ is chosen as the interior region of the two-dimensional unit disk centered at the origin, i.e.,
\[\mathcal{B}=\set{(\cos t,\sin t):t\in[0,2\pi]}.\]
The adopted wavenumber has the form $k_f=\frac{2\pi}{\lambda_f}$ for $f=1,2,\cdots,F$; here, $\lambda_f$ is the given wavelength that is equi-distributed between $\lambda_1=0.7$ and $\lambda_F=0.4$. Based on Theorem \ref{TheoremSymmetric}, the incident direction $\vec{\theta}_m$ is selected as
\[\vec{\theta}_m:=\bigg(\cos\frac{2(m-1)\pi}{M},\sin\frac{2(m-1)\pi}{M}\bigg),\quad m=1,2,\cdots,M,\]
for an even number $M$.

Motivated by \cite{K}, we choose two single cracks $\mathcal{C}_1$ and $\mathcal{C}_2$ as
\begin{align*}
  \mathcal{C}_1&=\set{\left(0.6t,\frac{1}{2}\cos\frac{t\pi}{2}+\frac{1}{5}\sin\frac{t\pi}{2}-\frac{1}{10}\cos\frac{3t\pi}{2}\right):t\in[-1,1]}\\
  \mathcal{C}_2&=\set{\left(1.5\sin\frac{(3t+4)\pi}{8}-1, 0.8\sin\frac{(3t+4)\pi}{4}\right):t\in[-1,1]},
\end{align*}
and inspired by \cite{PL3}, we choose multiple cracks $\mathcal{C}_{\mathrm{M}}$ as
\[\mathcal{C}_{\mathrm{ M}}=\{(t-0.2,-0.5t^2+0.5):t\in[-0.5,0.5]\}\cup\{(t+0.2,t^3+t^2-0.5):s\in[-0.5,0.5]\}.\]

Note that in order to show the robustness of the proposed algorithm, a white Gaussian noise with $15$ dB signal-to-noise ratio (SNR) is added to the unperturbed data.

\begin{figure}[!ht]
\begin{center}
\includegraphics[width=0.49\textwidth,keepaspectratio=true,angle=0]{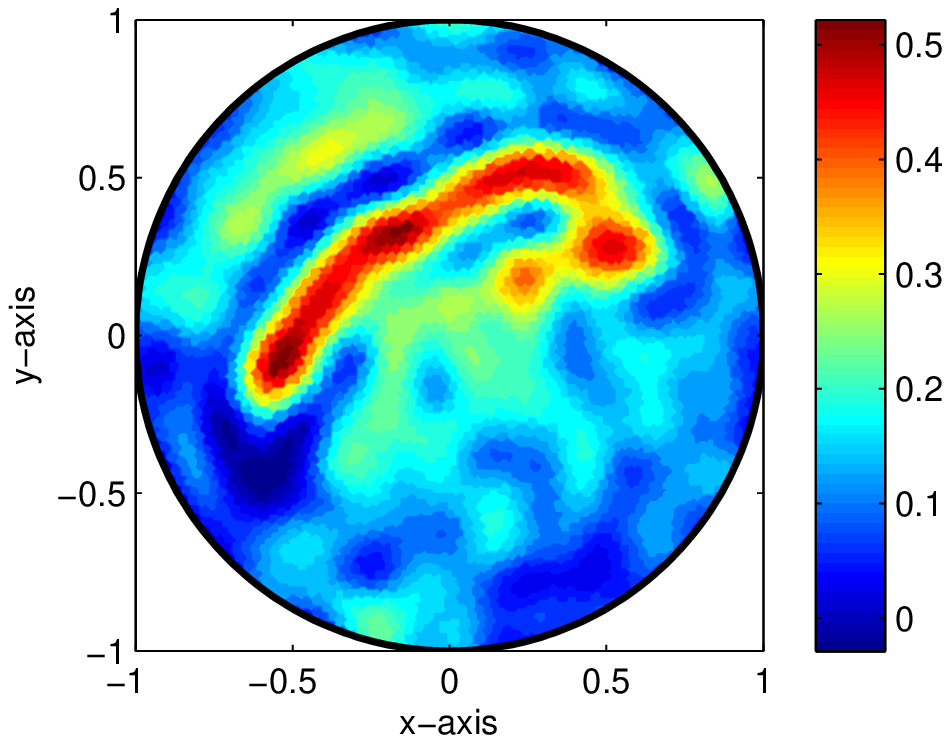}
\includegraphics[width=0.49\textwidth,keepaspectratio=true,angle=0]{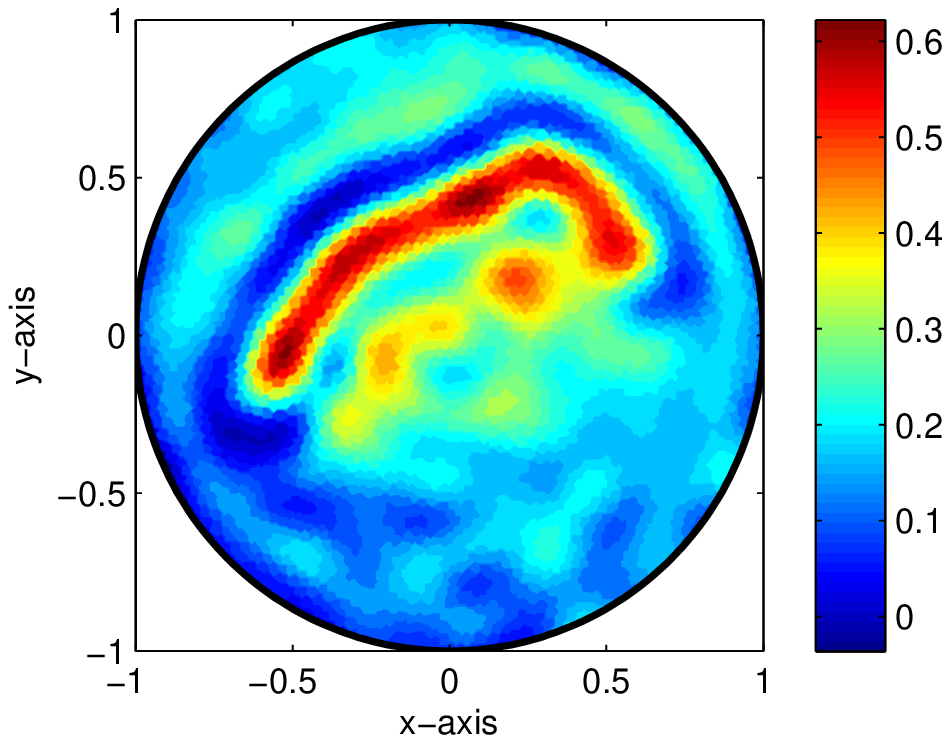}
\caption{\label{Gamma1}Map of $\mathbb{R}_{\mathrm{MTD}}(\vec{x}_s;F)$ with $F=16,M=4$ (left) and $F=16,M=16$ (right) for $\mathcal{C}_1$.}
\end{center}
\end{figure}

First, let us consider the imaging result of $\mathcal{C}_1$. The left-hand side of Figure \ref{Gamma1} shows the map of $\mathbb{R}_{\mathrm{MTD}}(\vec{x}_s;4)$ for $M=16$ different directions. In this result, because of its small magnitude at $(0.46,0.42)$, $\mathcal{C}_1$ appears divided. Moreover, owing to the large magnitude at certain points (for example, $(0.23,0.18)$), we cannot recognize its shape at this stage. The right-hand side of Figure \ref{Gamma1} is the map of $\mathbb{R}_{\mathrm{MTD}}(\vec{x}_s;16)$. The figure shows that there is undivided or whole $\mathcal{C}_1$; this means that we obtained a more accurate result than the previous one. However, because certain points have a large magnitude, we cannot determine the true shape of $\mathcal{C}_1$.

\begin{figure}[!ht]
\begin{center}
\includegraphics[width=0.49\textwidth,keepaspectratio=true,angle=0]{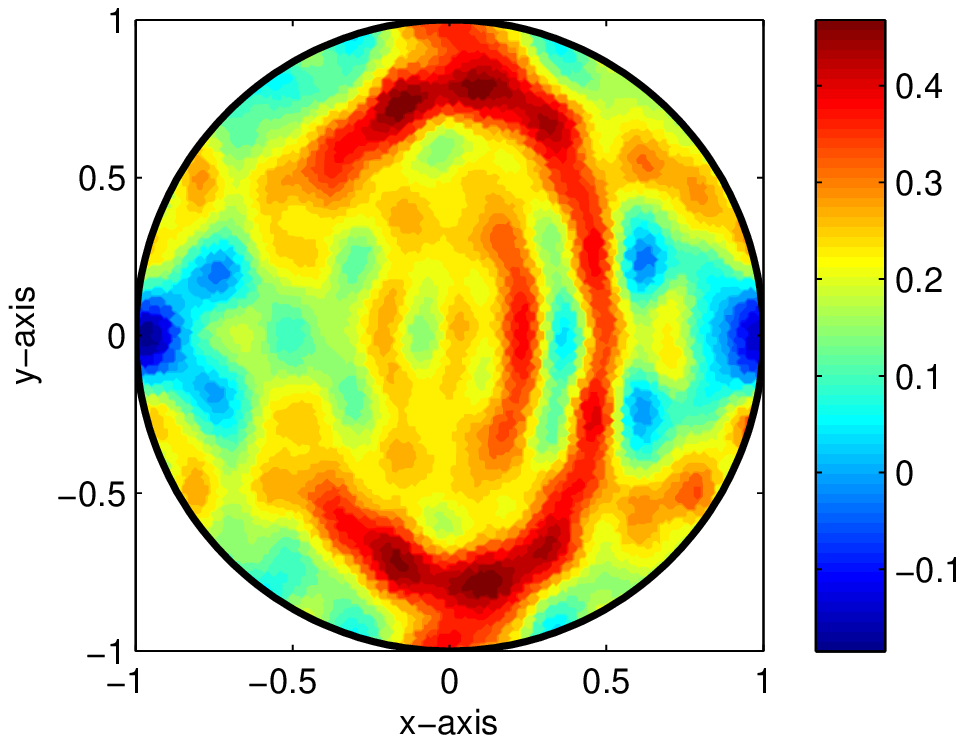}
\includegraphics[width=0.49\textwidth,keepaspectratio=true,angle=0]{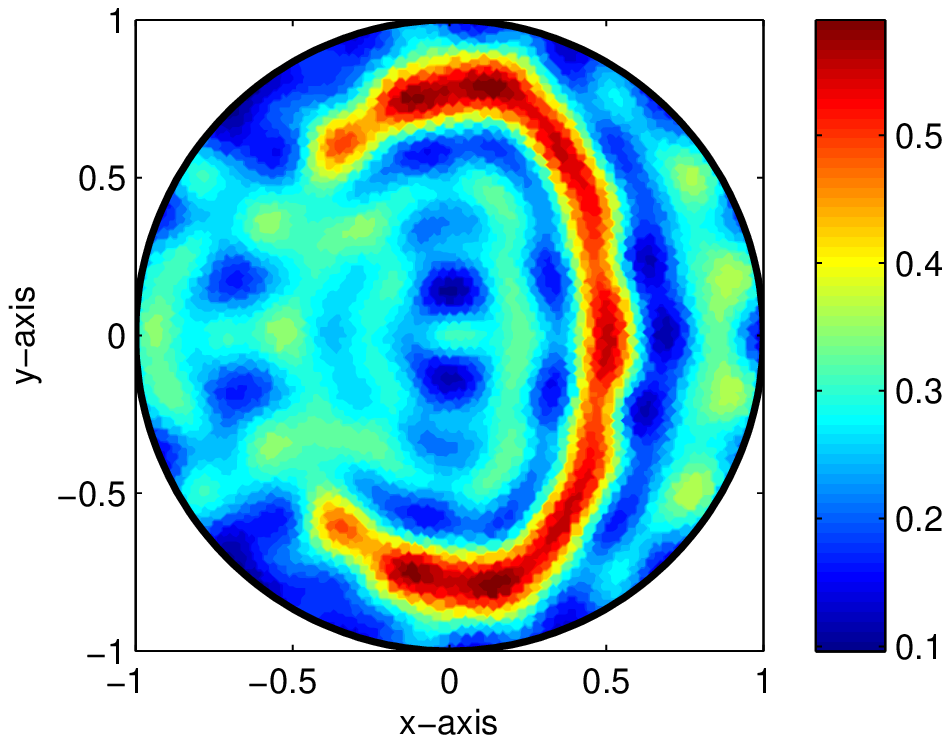}
\caption{\label{Gamma2}Map of $\mathbb{R}_{\mathrm{MTD}}(\vec{x}_s;F)$ with $F=16,M=4$ (left) and $F=16,M=16$ (right) for $\mathcal{C}_2$.}
\end{center}
\end{figure}

Figure \ref{Gamma2} shows the map of $\mathbb{R}_{\mathrm{MTD}}(\vec{x}_s;F)$ for crack $\mathcal{C}_2$. In contrast with the result of $\mathcal{C}_1$, it is difficult to identify the shape of the crack because of the great number of points with large magnitudes. Fortunately, similar to the results in Figure \ref{Gamma1}, map of $\mathbb{R}_{\mathrm{MTD}}(\vec{x}_s;16)$ yields a better image than that obtained by the map of $\mathbb{R}_{\mathrm{MTD}}(\vec{x}_s;4)$.

\begin{figure}[!ht]
\begin{center}
\includegraphics[width=0.49\textwidth,keepaspectratio=true,angle=0]{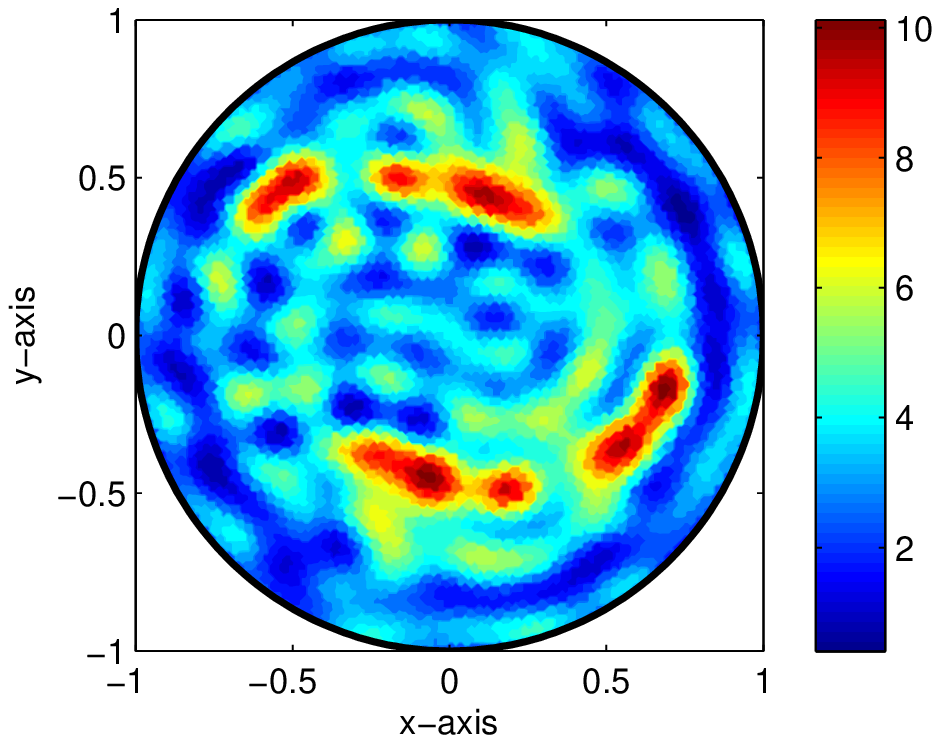}
\includegraphics[width=0.49\textwidth,keepaspectratio=true,angle=0]{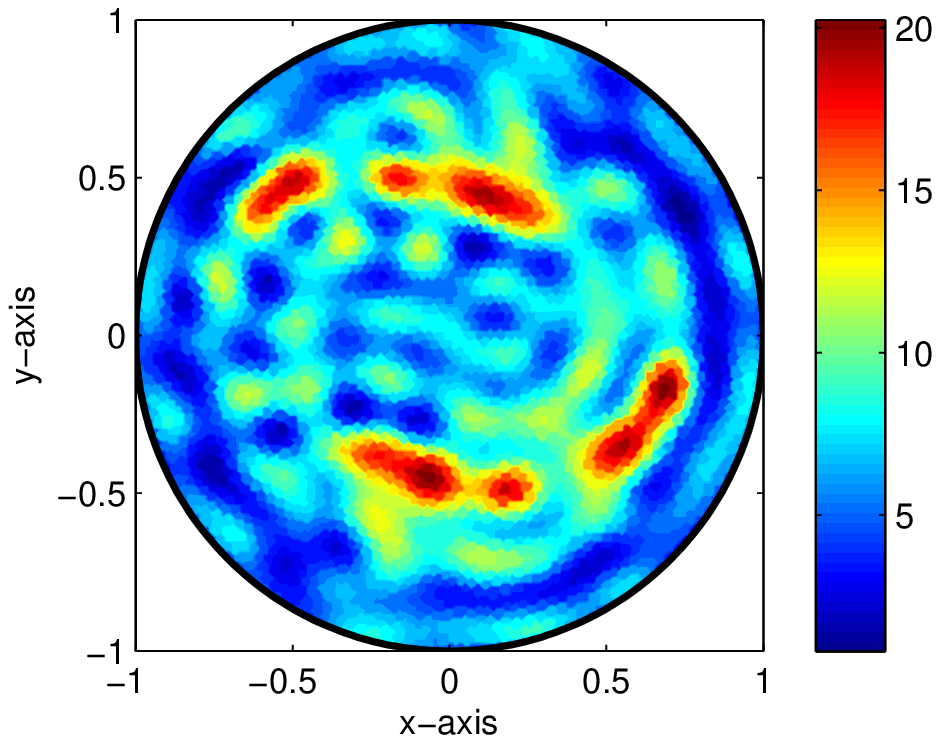}\\
\includegraphics[width=0.49\textwidth,keepaspectratio=true,angle=0]{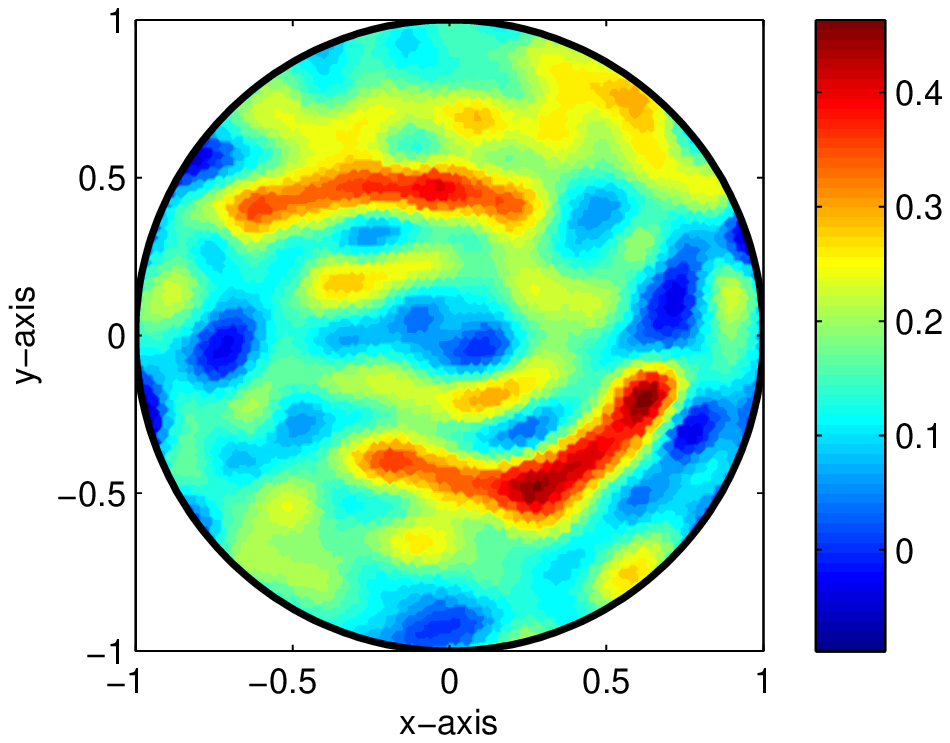}
\includegraphics[width=0.49\textwidth,keepaspectratio=true,angle=0]{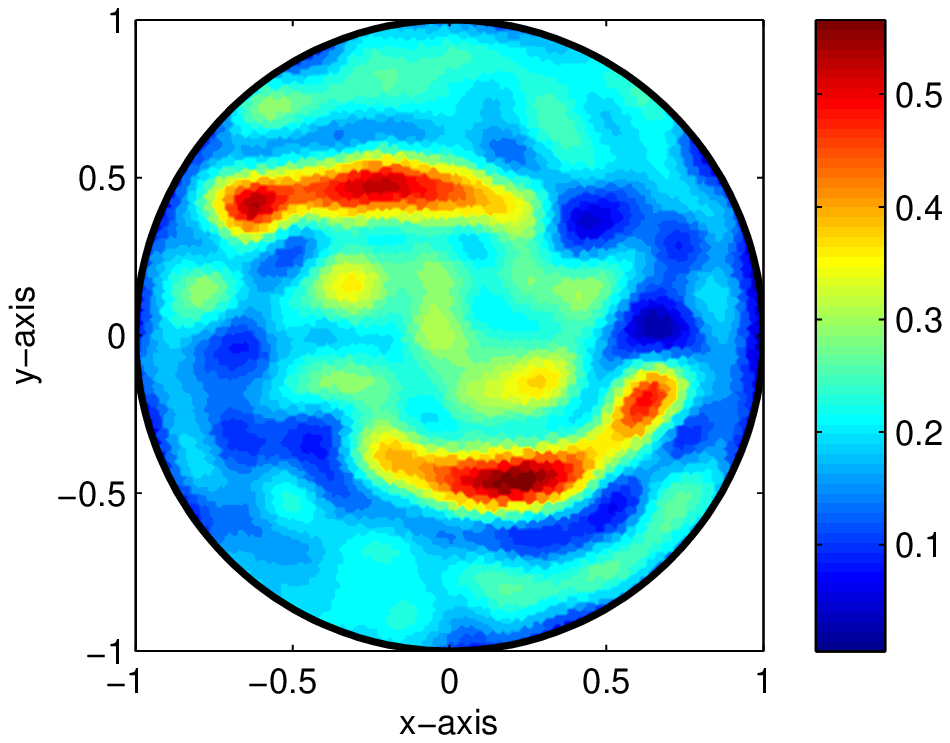}
\caption{\label{GammaM}Maps of $\mathbb{R}_{\mathrm{TD}}(\vec{x}_s;\frac{2\pi}{0.4})$ with $M=32$ (top, left) and $M=64$ (top, right), and maps of $\mathbb{R}_{\mathrm{MTD}}(\vec{x}_s;F)$ with $F=16,M=4$ (bottom, left) and $F=16,M=16$ (bottom, right) for $\mathcal{C}_{\mathrm{M}}$.}
\end{center}
\end{figure}

The application of multiple cracks $\mathcal{C}_{\mathrm{M}}$ is shown in Figure \ref{GammaM}. This figure shows the maps of $\mathbb{R}_{\mathrm{TD}}(\vec{x}_s;\frac{2\pi}{0.4})$ and $\mathbb{R}_{\mathrm{MTD}}(\vec{x}_s;F)$. Throughout these results, we can observe that the map of $\mathbb{R}_{\mathrm{TD}}(\vec{x}_s;k)$ yields a poor result even if the number of incident directions $M$ is large; conversely, the multi-frequency imaging function $\mathbb{R}_{\mathrm{MTD}}(\vec{x}_s;F)$ yields an accurate result even if $M$ is small.

\begin{figure}[!ht]
\begin{center}
\includegraphics[width=0.49\textwidth,keepaspectratio=true,angle=0]{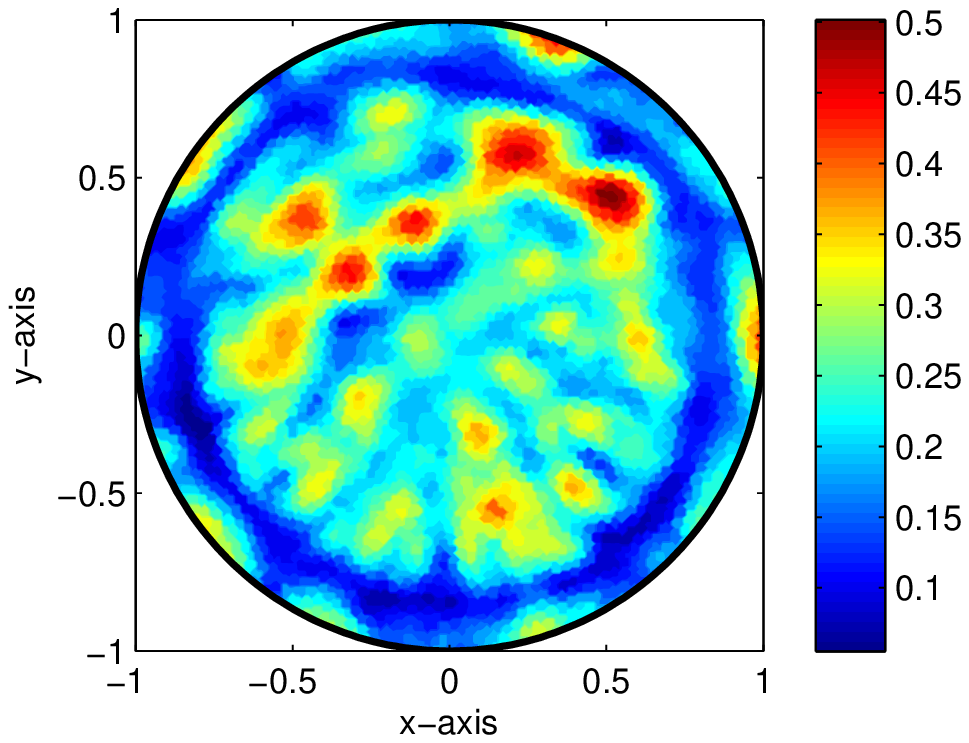}
\includegraphics[width=0.49\textwidth,keepaspectratio=true,angle=0]{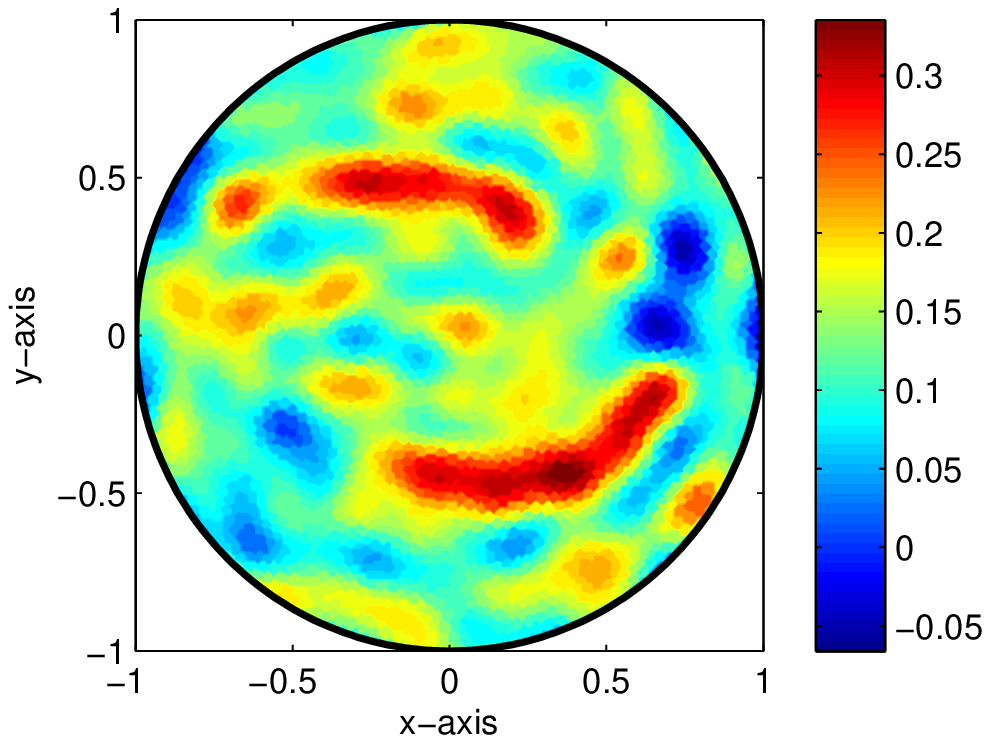}
\caption{\label{GammaNonSymmetric1}Map of $\mathbb{R}_{\mathrm{MTD}}(\vec{x}_s;F)$ with $F=16,M=5$ for $\mathcal{C}_1$ (left) and $\mathcal{C}_{\mathrm{M}}$ (right).}
\end{center}
\end{figure}

\begin{figure}[!ht]
\begin{center}
\includegraphics[width=0.49\textwidth,keepaspectratio=true,angle=0]{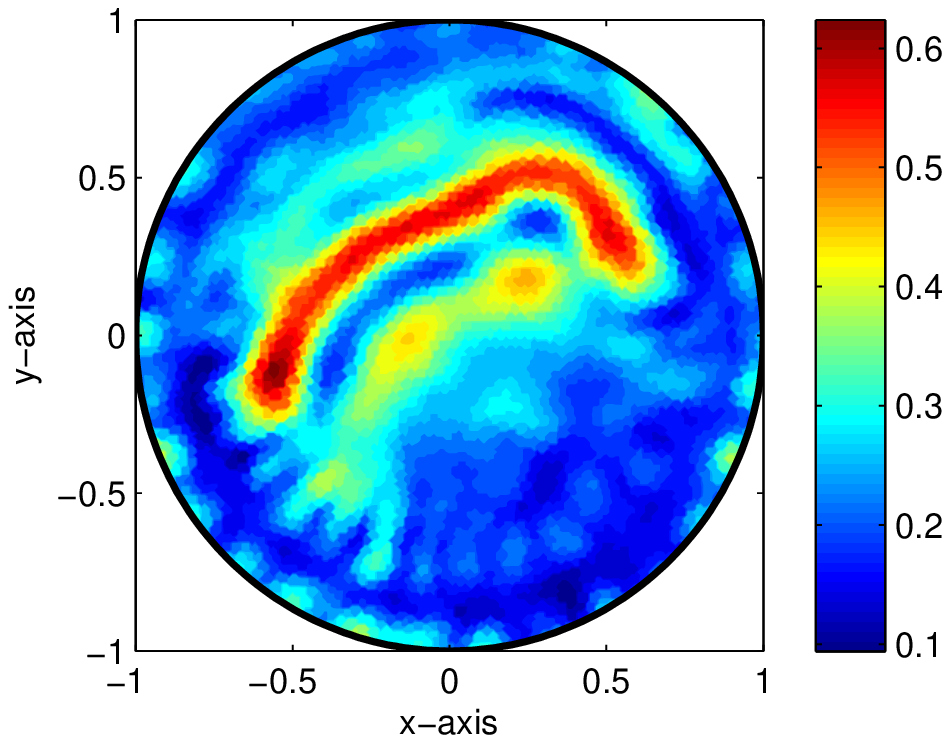}
\includegraphics[width=0.49\textwidth,keepaspectratio=true,angle=0]{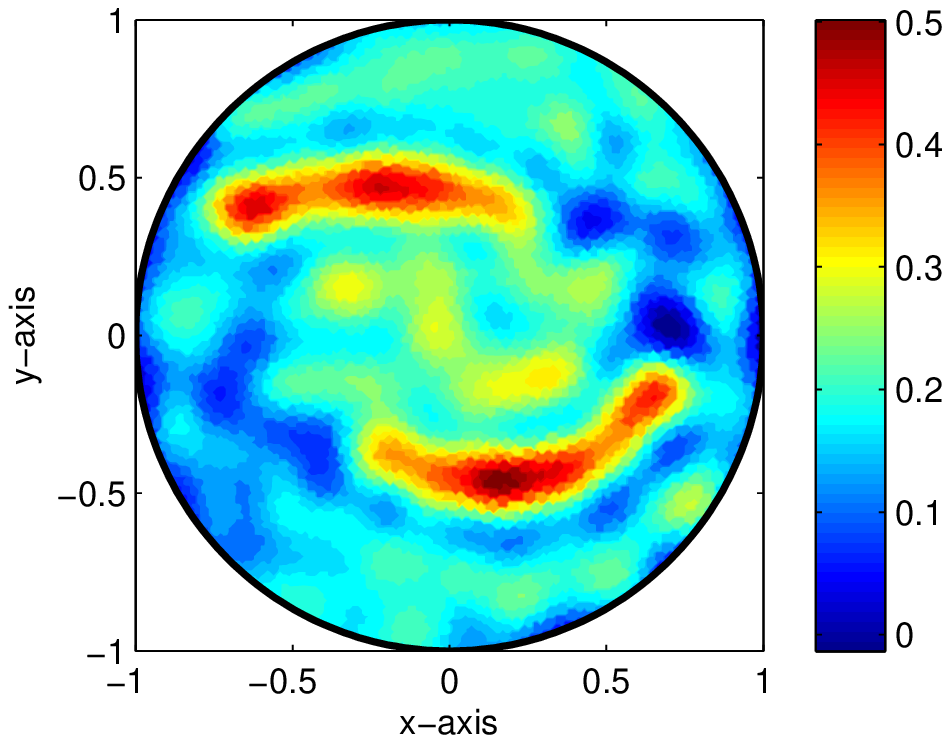}
\caption{\label{GammaNonSymmetric2}Map of $\mathbb{R}_{\mathrm{MTD}}(\vec{x}_s;F)$ with $F=16,M=15$ for $\mathcal{C}_1$ (left) and $\mathcal{C}_{\mathrm{M}}$ (right).}
\end{center}
\end{figure}

In Figure \ref{GammaNonSymmetric1}, maps of $\mathbb{R}_{\mathrm{MTD}}(\vec{x}_s;16)$ with $M=5$ number of incident directions are displayed for cracks $\mathcal{C}_1$ and $\mathcal{C}_{\mathrm{M}}$ in order to verify the \ref{P2Remark} of Remark \ref{Remark1}. By comparing Figures \ref{Gamma1} and \ref{GammaM}, it is difficult to discern the shape of true crack(s) because of the appearance of unforseen replicas with large magnitudes. Note that if $M$ is an odd number but is sufficiently large enough, the map of $\mathbb{R}_{\mathrm{MTD}}(\vec{x}_s;F)$ yield a good result, refer to Figure \ref{GammaNonSymmetric2}.

\begin{figure}[!ht]
\begin{center}
\includegraphics[width=0.49\textwidth,keepaspectratio=true,angle=0]{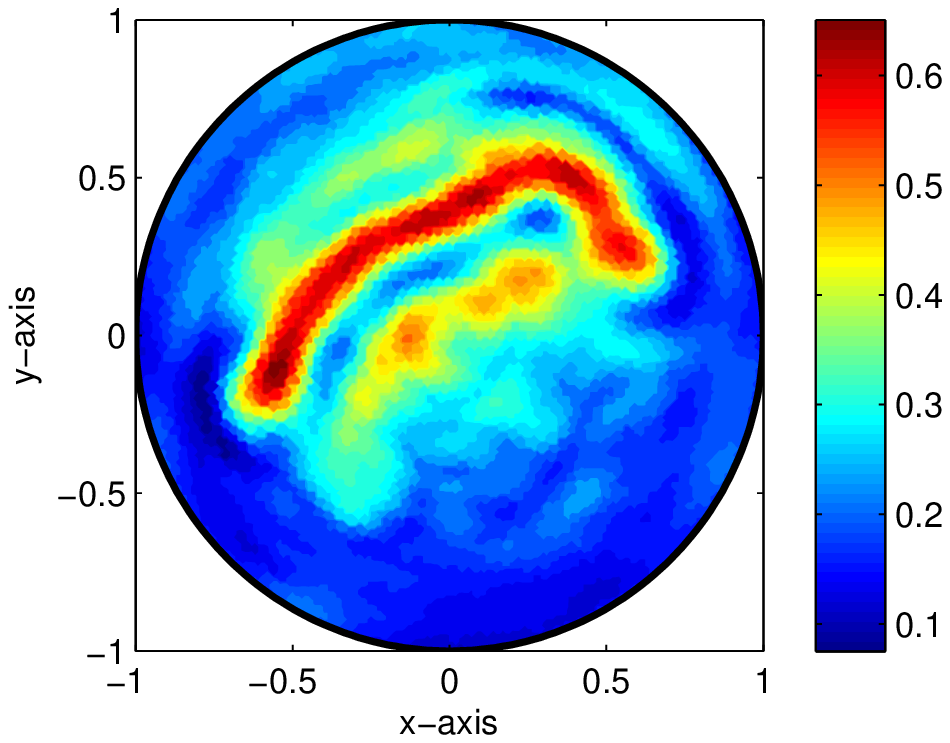}
\includegraphics[width=0.49\textwidth,keepaspectratio=true,angle=0]{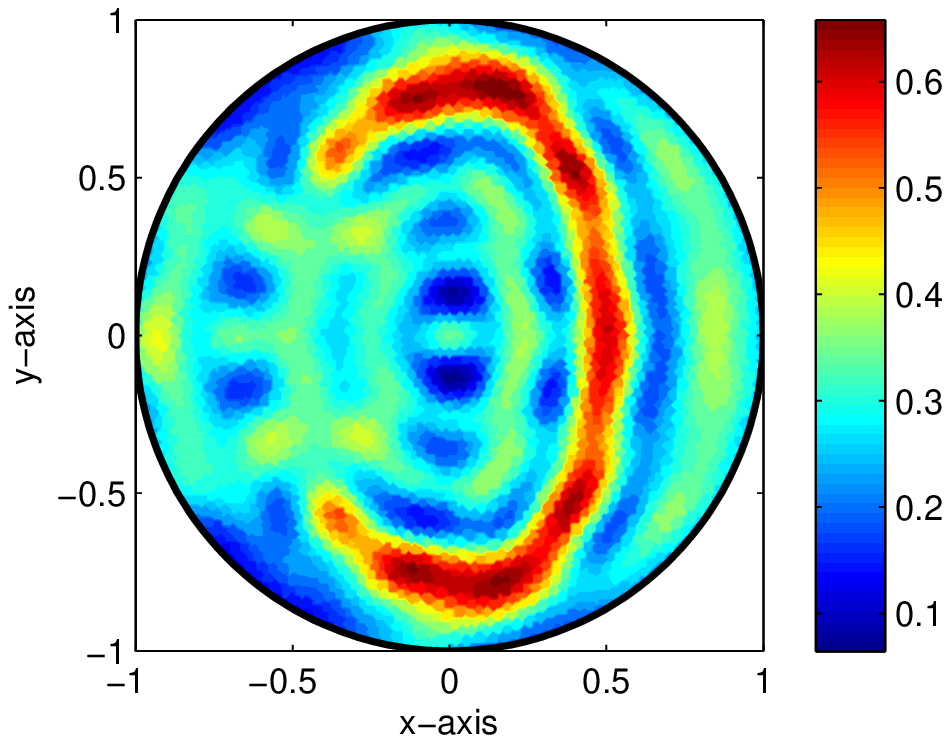}
\caption{\label{Gamma12-6416}Map of $\mathbb{R}_{\mathrm{MTD}}(\vec{x}_s;F)$ with $F=16,M=64$ for $\mathcal{C}_1$ (left) and $\mathcal{C}_2$ (right).}
\end{center}
\end{figure}

\begin{figure}[!ht]
\begin{center}
\includegraphics[width=0.49\textwidth,keepaspectratio=true,angle=0]{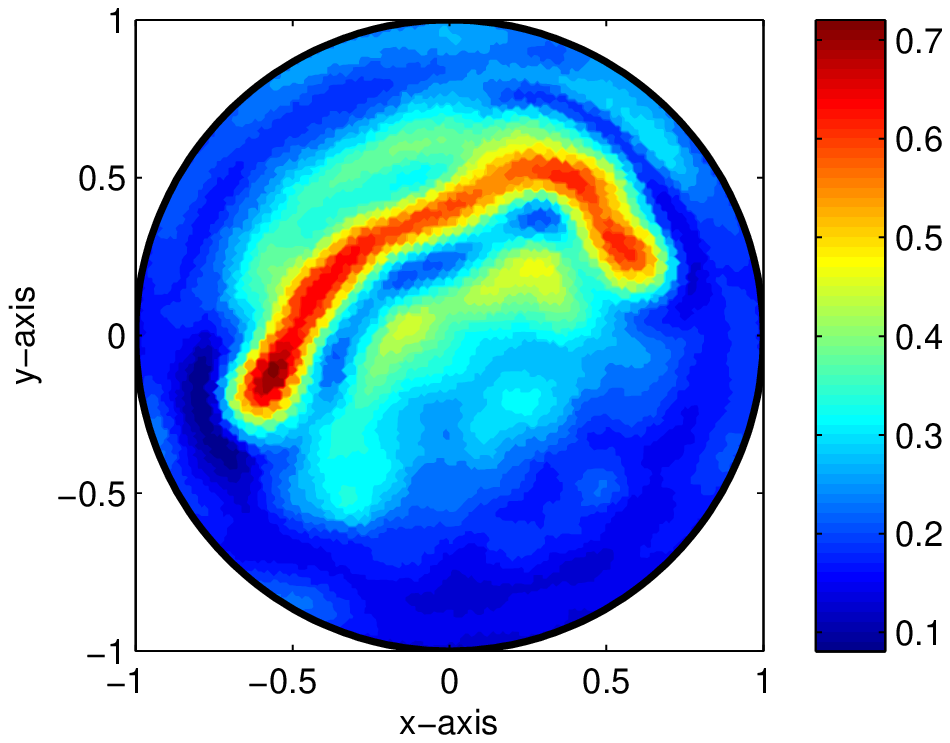}
\includegraphics[width=0.49\textwidth,keepaspectratio=true,angle=0]{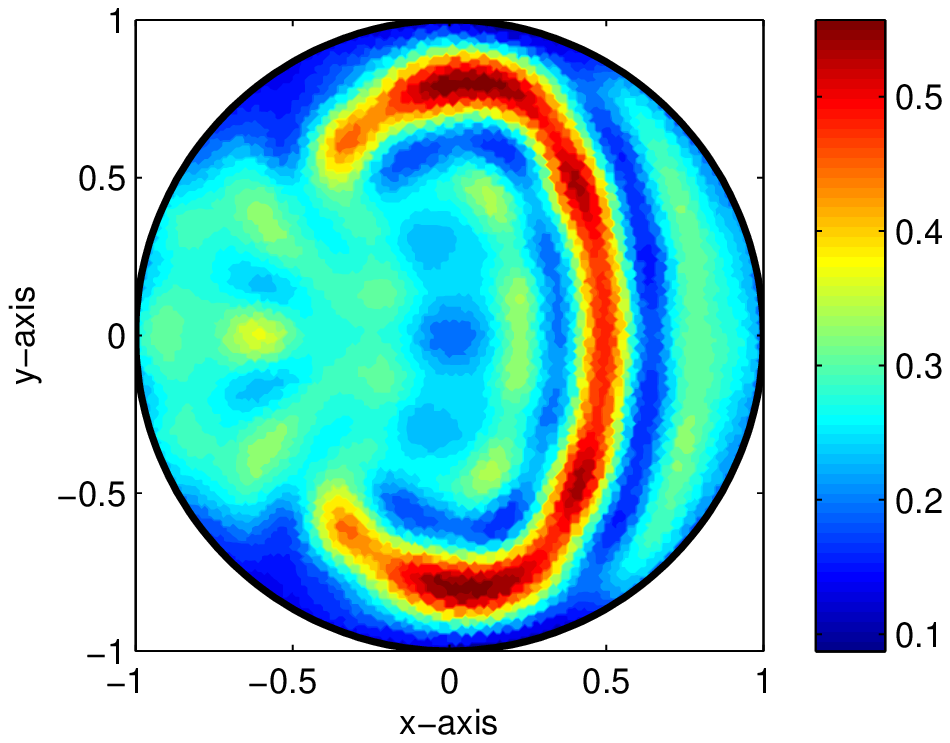}
\caption{\label{Gamma12-6464}Map of $\mathbb{R}_{\mathrm{MTD}}(\vec{x}_s;F)$ with $F=64,M=64$ for $\mathcal{C}_1$ (left) and $\mathcal{C}_2$ (right).}
\end{center}
\end{figure}

\begin{figure}[!ht]
\begin{center}
\includegraphics[width=0.49\textwidth,keepaspectratio=true,angle=0]{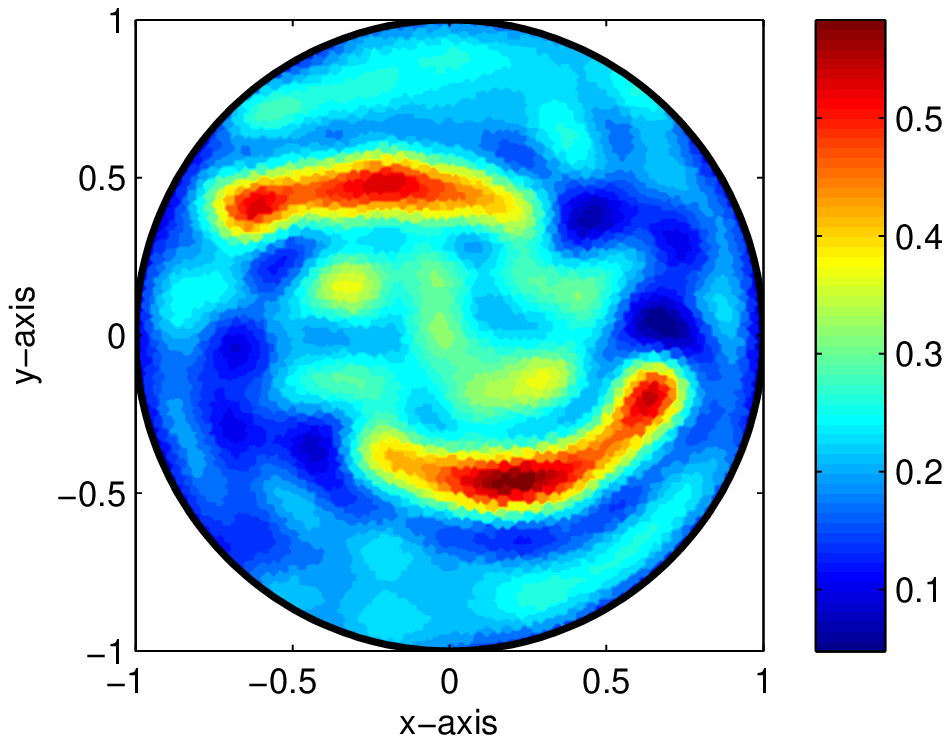}
\includegraphics[width=0.49\textwidth,keepaspectratio=true,angle=0]{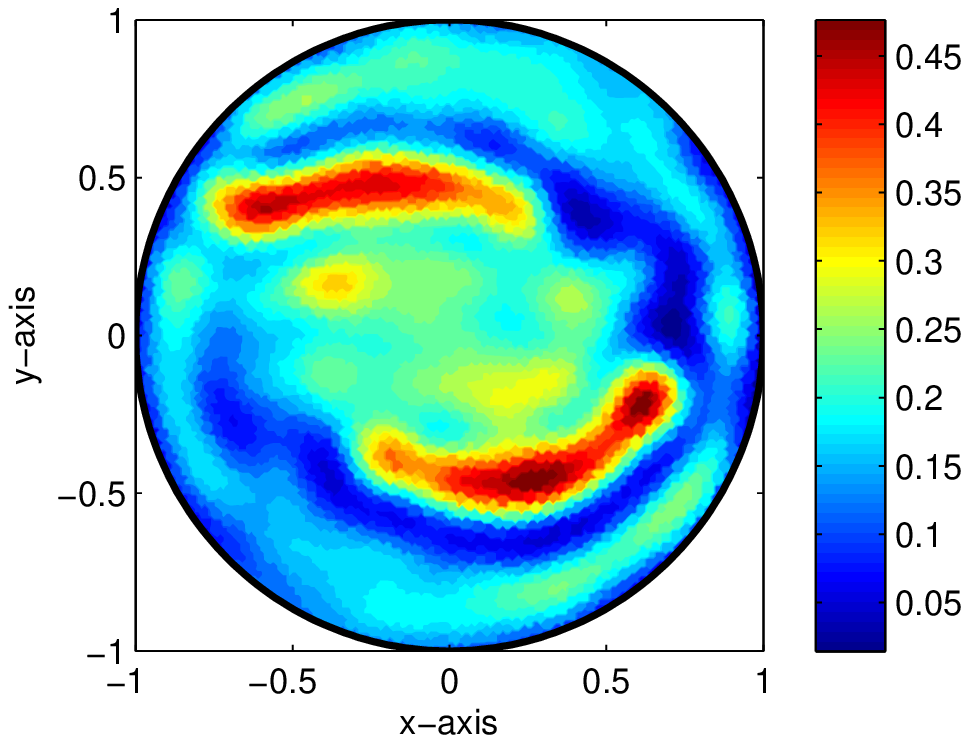}
\caption{\label{GammaM-64}Map of $\mathbb{R}_{\mathrm{MTD}}(\vec{x}_s;F)$ with $F=16,M=64$ (left) and $F=64,M=64$ (right) for $\mathcal{C}_{\mathrm{M}}$.}
\end{center}
\end{figure}

At this moment, we increase the number of applied frequencies $F$ and incident directions $M$. Figures \ref{Gamma12-6416}, \ref{Gamma12-6464}, and \ref{GammaM-64} depict the map of $\mathbb{R}_{\mathrm{MTD}}(\vec{x}_s;F)$ when $F=64$ and $M=16$ or $M=64$. As we expected in \ref{P3Remark} of Remark \ref{Remark1}, the results yielded by $\mathbb{R}_{\mathrm{MTD}}(\vec{x}_s;F)$ improve with increasing number of $F$ and/or $M$.

\begin{figure}[!ht]
\begin{center}
\includegraphics[width=0.49\textwidth,keepaspectratio=true,angle=0]{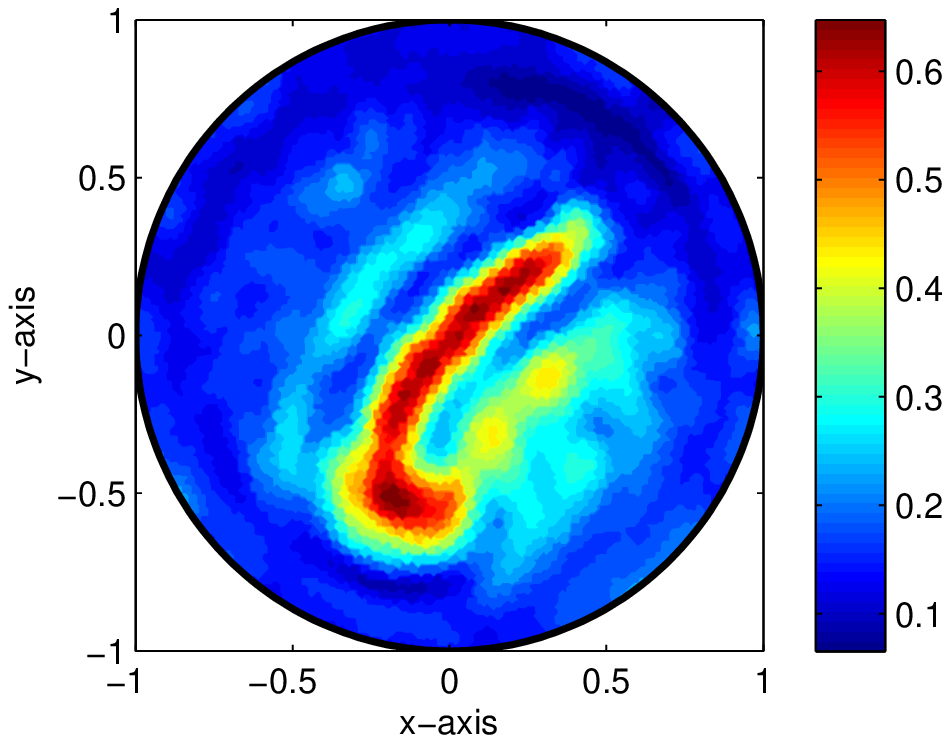}
\includegraphics[width=0.49\textwidth,keepaspectratio=true,angle=0]{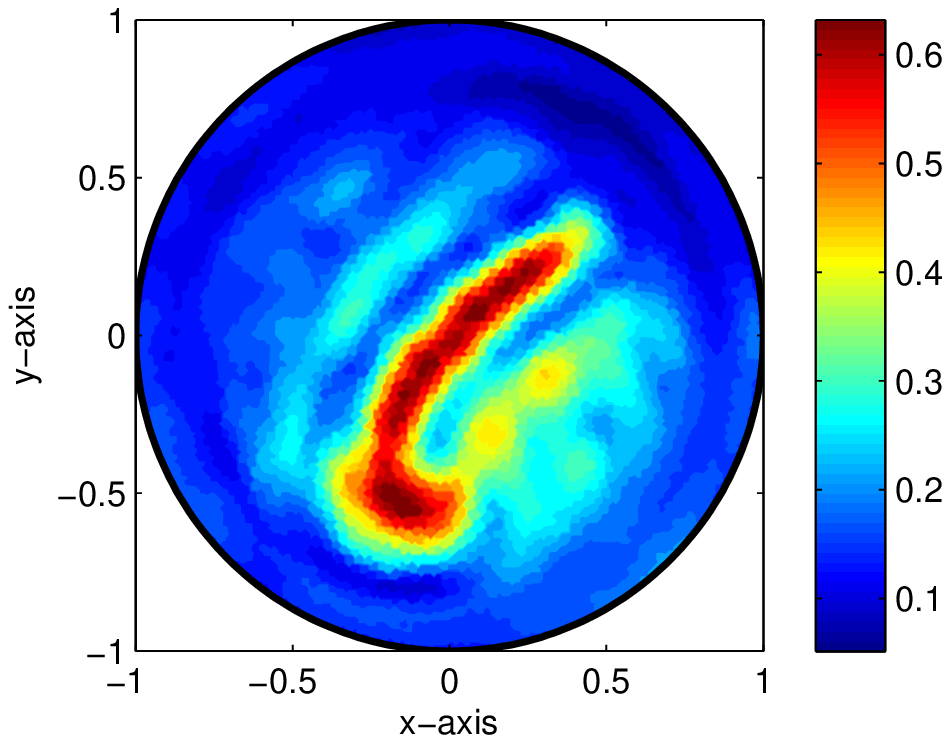}\\
\includegraphics[width=0.49\textwidth,keepaspectratio=true,angle=0]{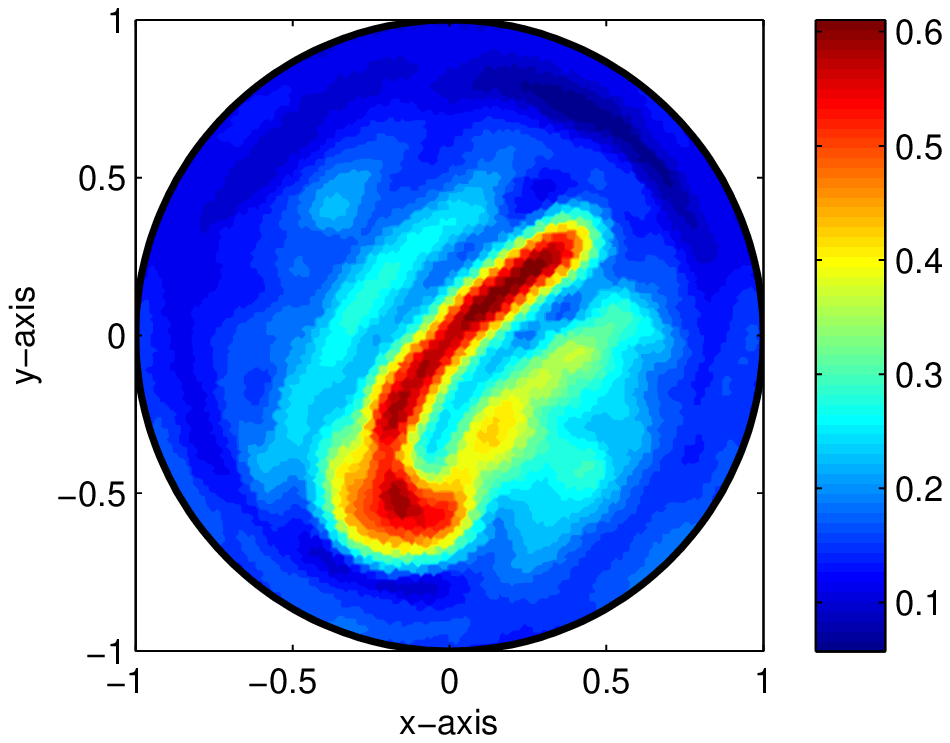}
\includegraphics[width=0.49\textwidth,keepaspectratio=true,angle=0]{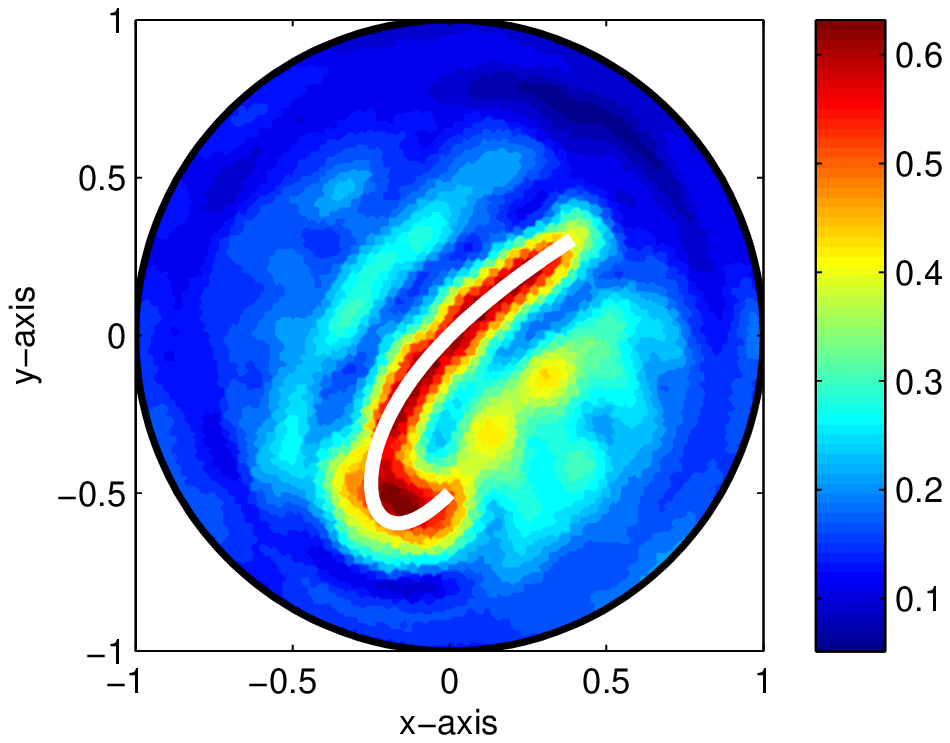}
\caption{\label{GammaA}Map of $\mathbb{R}_{\mathrm{MTD}}(\vec{x}_s;F)$ for $\mathcal{C}_3$ with $F=16,M=16$ (top, left), $F=16,M=64$ (top, right), $F=64,M=64$ (bottom, left) and true shape (bottom, right).}
\end{center}
\end{figure}

\begin{figure}[!ht]
\begin{center}
\includegraphics[width=0.49\textwidth,keepaspectratio=true,angle=0]{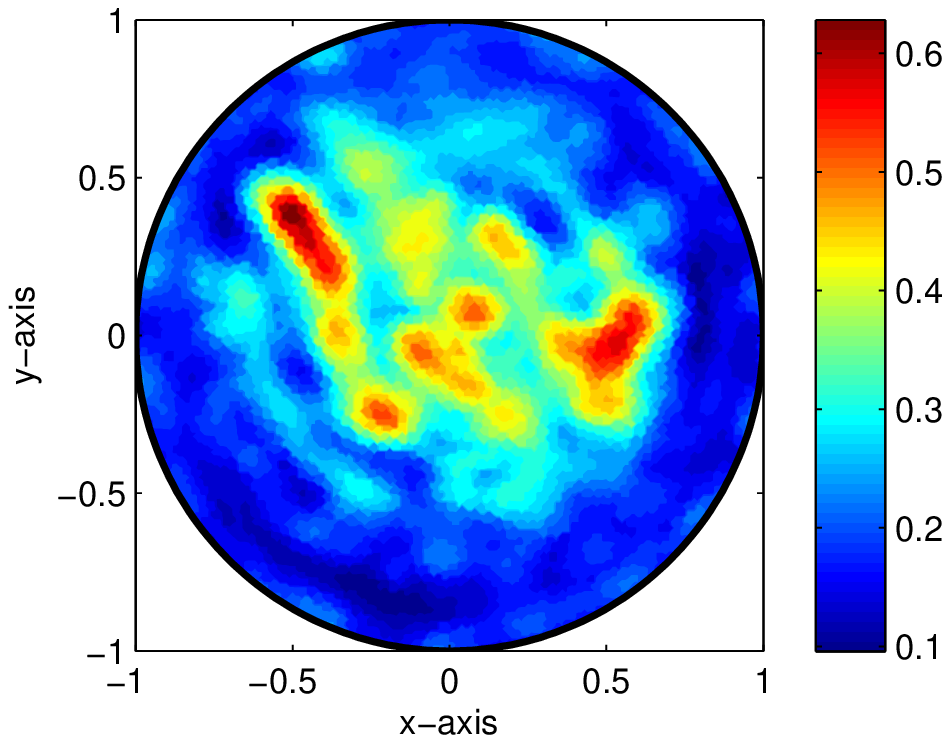}
\includegraphics[width=0.49\textwidth,keepaspectratio=true,angle=0]{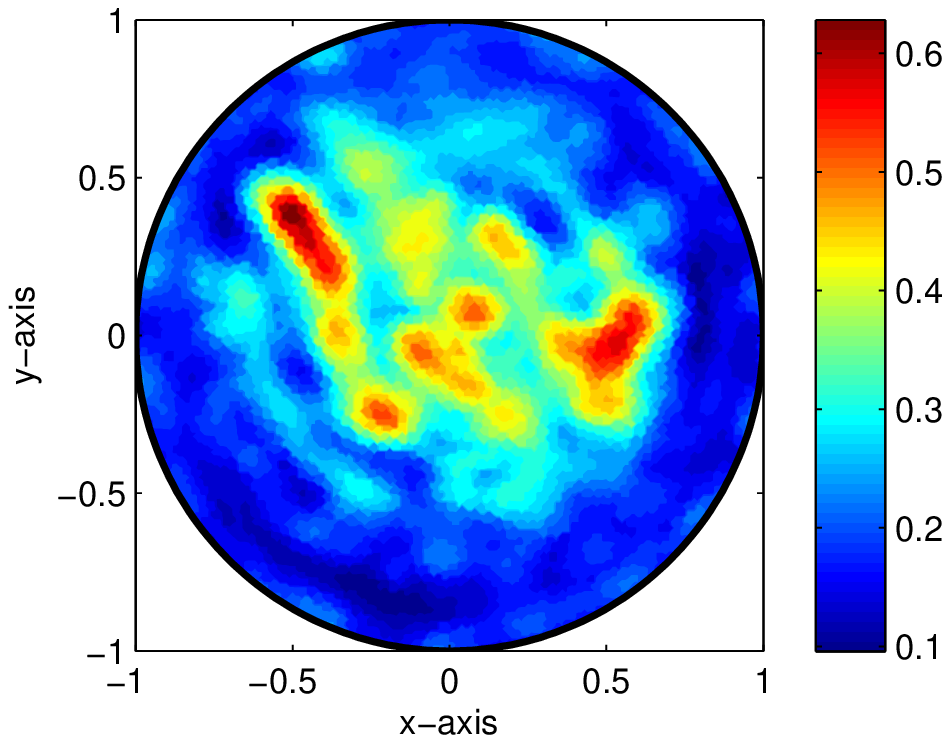}\\
\includegraphics[width=0.49\textwidth,keepaspectratio=true,angle=0]{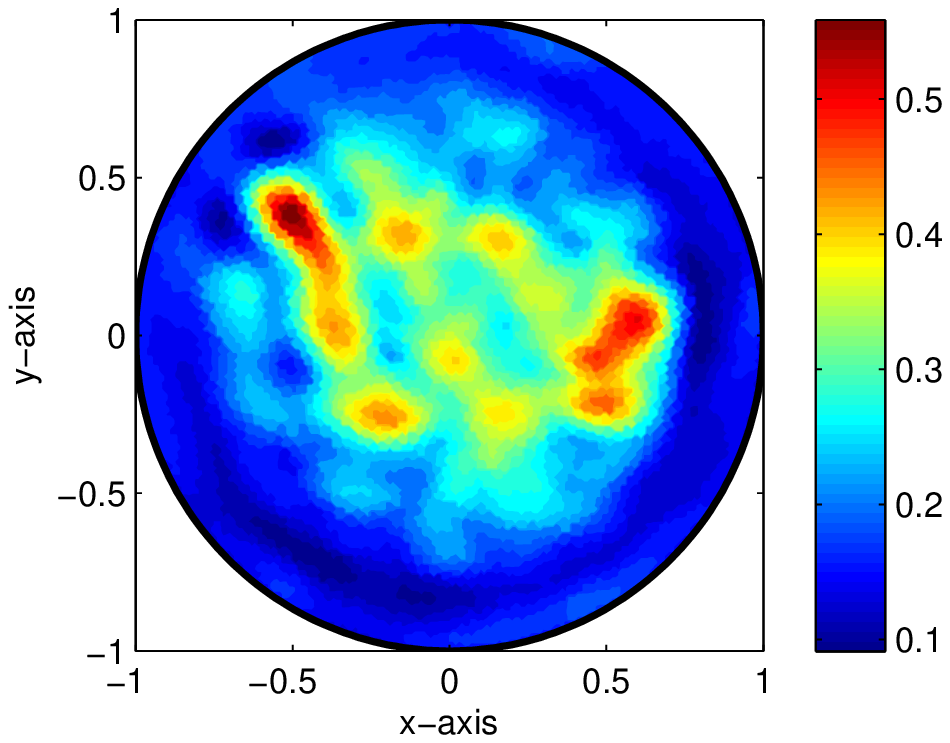}
\includegraphics[width=0.49\textwidth,keepaspectratio=true,angle=0]{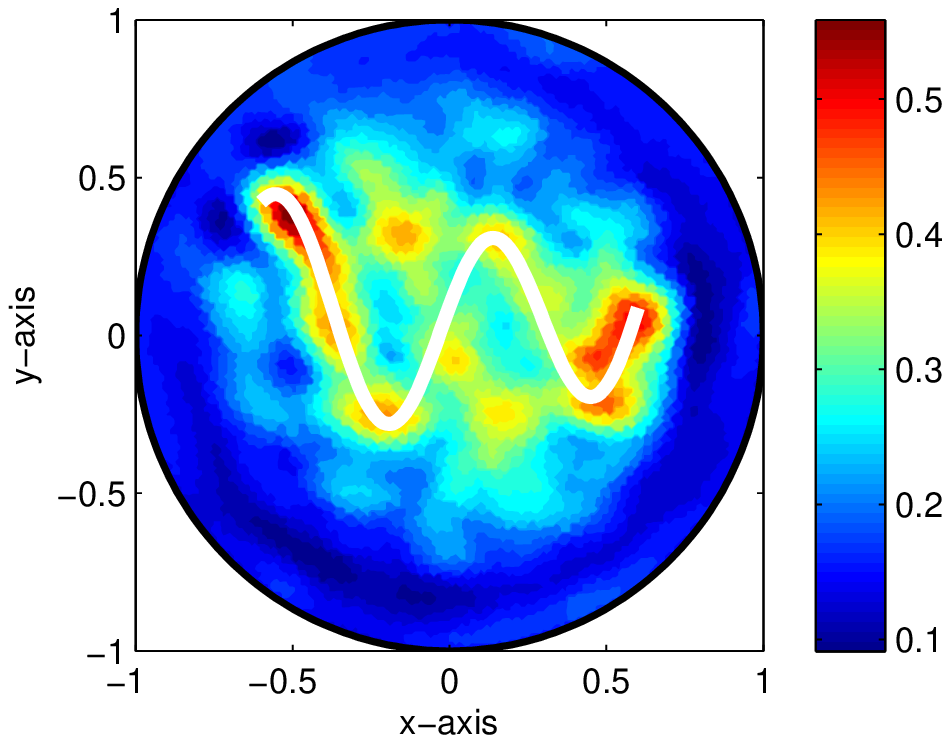}
\caption{\label{GammaB}Same as Figure \ref{GammaA} except the crack is $\mathcal{C}_4$.}
\end{center}
\end{figure}

\begin{Rem}[Limitation of the multi-frequency topological derivative]
  Although multi-frequency topological derivative imaging yields very good imaging results for not only single but also multiple cracks, it still has the following limitations.
  \begin{enumerate}
    \item Let us consider the imaging of a crack of a large curvature near one of the crack tips. To illustrate this, we choose a crack from \cite{KS}:
        \[\mathcal{C}_3=\set{(t+t^2,0.5t^4+t):-1\leq t\leq3}.\]

        From the results in Figure \ref{GammaA}, the imaging results yielded by $\mathbb{R}_{\mathrm{MTD}}(\vec{x}_s;F)$ are satisfactory. However, at the point of the large curvature, the map of $\mathbb{R}_{\mathrm{MTD}}(\vec{x}_s;F)$ reconstructs a slightly different shape of the crack.
    \item We apply the proposed algorithm for imaging an oscillation crack represented as follows:
        \[\mathcal{C}_4=\set{(t,0.5t^2+0.1\sin(3\pi(t+0.7))):-0.6\leq t\leq0.6}.\]

        Figure \ref{GammaB} shows the map of $\mathbb{R}_{\mathrm{MTD}}(\vec{x}_s;F)$. The obtained result by the proposed algorithm does not improve even when the number of $F$ and $M$ are increased. The proposed algorithm requires further optiization.
  \end{enumerate}
\end{Rem}

\section{Concluding remarks}\label{Sec5}
We improved the traditional topological derivative concept to determine location and reconstruct the shape of arbitrary-shaped curve-like perfectly conducting cracks. For this purpose, we designed a reconstruction algorithm based on the multi-frequency topological derivative, analyzed its structure, and investigated its properties. We confirmed that the multi-frequency topological derivative contains singularity on the cracks, and therefore it reaches its maximum value at the location of the cracks. Moreover, we demonstrated that why ghost replicas appeared in the neighborhood of cracks and why the configuration of symmetric incident directions yields better results than the non-symmetric configuration. Numerical simulations under various situations depicted both the benefits and limitations of the proposed algorithm. Because the algorithm requires only a one-step iteration procedure, it is very fast in terms of shape identification but does not yield the complete shape of cracks. However, by adopting the obtained result as a starting point of a Newton-type based algorithm \cite{K}, it is expected that complete shape reconstruction can be successfully accomplished.

This paper deals with the reconstruction of perfectly conducting cracks with Dirichlet boundary condition. Accordingly, the extension of this research to the crack with Neumann boundary condition will be an interesting work. We believe that applying higher-order terms in the asymptotic expansion formula \cite{AK} give a higher order topological derivative \cite{B}. The calculation and analysis of higher-order topological derivative will be a valuable research topic. According to the Statistical Hypothesis Testing \cite{AGKPS}, multi-frequencies are expected to enhance the imaging performance. The corresponding work will include a careful stability and resolution analysis of the multi-frequency topological derivative.

\end{document}